# Exploring the impacts of conformer selection methods on ion mobility collision cross section predictions

Felicity F. Nielson, Sean M. Colby, Dennis G. Thomas, Ryan S. Renslow*, Thomas O. Metz*

Biological Sciences Division, Pacific Northwest National Laboratory, Richland, WA, USA

*Corresponding authors: ryan.renslow@pnnl.gov; thomas.metz@pnnl.gov

## ABSTRACT

The prediction of structure dependent molecular properties, such as collision cross sections as measured using ion mobility spectrometry, are crucially dependent on the selection of the correct population of molecular conformers. Here, we report an in-depth evaluation of multiple conformation selection techniques, including simple averaging, Boltzmann weighting, lowest energy selection, low energy threshold reductions, and similarity reduction. Generating 50,000 conformers each for 18 molecules, we used the *In Silico* Chemical Library Engine (ISiCLE) to calculate the collision cross sections for the entire dataset. First, we employed Monte Carlo simulations to understand the variability between conformer structures as generated using simulated annealing. Then we employed Monte Carlo simulations to the aforementioned conformer selection techniques applied on the simulated molecular property—the ion mobility collision cross section. Based on our analyses, we found Boltzmann weighting to be a good tradeoff between precision and theoretical accuracy. Combining multiple techniques revealed that energy thresholds and root-mean-squared deviation-based similarity reductions can save considerable computational expense while maintaining property prediction accuracy. Molecular dynamic conformer generation tools like AMBER can continue to generate new lowest energy conformers even after tens of thousands of generations, decreasing precision between runs. This reduced precision can be ameliorated and theoretical accuracy increased by running density functional theory geometry optimization on carefully selected conformers.

## INTRODUCTION

The identification and quantification of small molecules – metabolomics – has a broad range of applications, from forensics [1-3] to human health and disease [3-5], soil microbiology [6-9] and materials science [10,11]. The current gold standard methods for identifying small molecules in complex samples rely on comparing experimental data (i.e., observed "features") to libraries derived from pure chemicals analyzed using the same experimental platform. Such reference materials are limited in availability and can be costly to acquire en masse, especially at high purity, and can require significant time to process and analyze. The vast majority of molecules in the universe are yet undiscovered, and even of those that are known, most are not readily available for purchase [12-15]. It has therefore become crucial to develop computational methods for building reliable libraries of predicted molecular properties that are validated against empirical experiments in order to reduce reliance on authentic reference materials.

Many groups have developed methods for predicting chemical properties measured in several identifications platforms including quantitative structure-retention relationship and machine learning models to predict liquid chromatography retention times [16,17], combinatorial approaches to predict MS/MS fragmentation patterns [18-20], quantum chemical calculations and artificial neural networks for NMR chemical shift predictions [21-24], and classical scattering and machine/deep learning to predict CCS for IMS [25-30]. Similarly, our group recently developed the *In Silico* Chemical Library Engine (ISiCLE), which is an automated workflow for molecular property calculation. It has shown preliminary success for calculating collision cross sections (CCS) and NMR chemical shifts [31,32].

Chemical properties and molecular behavior are a consequence of inter- and intra-molecular forces, as governed primarily by the electron distribution surrounding the constituent nuclei. Therefore, the conformer, or specific 3D structure of a molecule that otherwise has the same atoms and bonds, has significant impact on the outcome of chemical interactions. Many chemical properties (e.g. CCS and NMR chemical shifts) are highly sensitive to the underlying conformational populations, and consequently, nearly all of the computational approaches listed above require the initial step of generating conformers. Suitable conformer(s) must be chosen for accurate *in silico* molecular simulations or the results of chemical property predictions may be open to significant error.

To date, there has not been a clear study that has evaluated tradeoffs between various conformer sampling techniques—including analysis of the appropriate number to be used or the best method of selection. In this study, we explore several methods for conformer selection to assess the impact these approaches have on the prediction of the molecular property collision cross section (CCS), as measured using ion mobility spectrometry (IMS). In IMS, a sample of molecule(s) is ionized (e.g., via electrospray ionization) and then propelled by an electric field through a drift region populated by a neutral buffer gas (commonly nitrogen or helium). The momentum transfer and chemical interactions between the molecular ions (adducts) and the buffer gas result in changing the net drift velocity of ion packets, leading to the separation of molecular adducts, including adducts for isomers [33] and isotopologues [34]. The measured arrival time of the ions at the end of the drift region can be used to calculate the CCS. As CCS is a property of both the ion and the buffer gas, different buffer gases will yield different CCS values for the same ion.

During an IMS separation, a single molecular adduct is not associated with a single spike in arrival time, but rather a distribution of arrival times as seen in **Fig. 1a**. This distribution is due to ion packet diffusion and multiple, interconverting conformers existing simultaneously within each packet. Commonly, the arrival time associated with the peak apex of the arrival time distribution is used to calculate a single experimental CCS for each molecular adduct. *In silico* predictions of CCS that are based on molecular structure (as opposed to 3D structure-naïve approaches [29,30]) therefore ideally choose a conformer or group of conformers that will result in CCS that are as close as possible to the experimental values represented by the arrival time peak maxima. The more accurate the predicted CCS, the more useful they will be for identification libraries and for reducing molecular identification false positive rates.

In this study, we used Monte Carlo, energy cutoff, and similarity downselection methods, as well as a variety of averaging methods, to explore how varying the number and type of conformers considered in a modified ISiCLE pipeline relate to final CCS predictions. We used a benchmark set of molecules with experimentally determined CCS, spanning various size and molecular flexibility, to compare the different methods. We found the average variability between conformer geometries, and therefore the number of conformers needed to represent conformational space, was positively correlated with molecular properties, such as having a high chain bond or atom count and rotatable bond count (high degrees of freedom), and negatively correlated with $2^{nd}$ acidic pKa sites.

Similar to current literature methods, for calculating CCS we found Boltzmann weighting yielded the best result among evaluated averaging techniques. We also found the conformers generated from MD simulations using AMBER insufficiently covered the low energy region of conformational space, leading to lower precision between simulations. DFT geometry optimization helps resolve this sparsity issue, which may also be present with other conformer generation tools. While this study focuses on the effect of conformer selection on CCS, we believe the results and the methods of evaluation can be generalized to other molecular modeling applications.

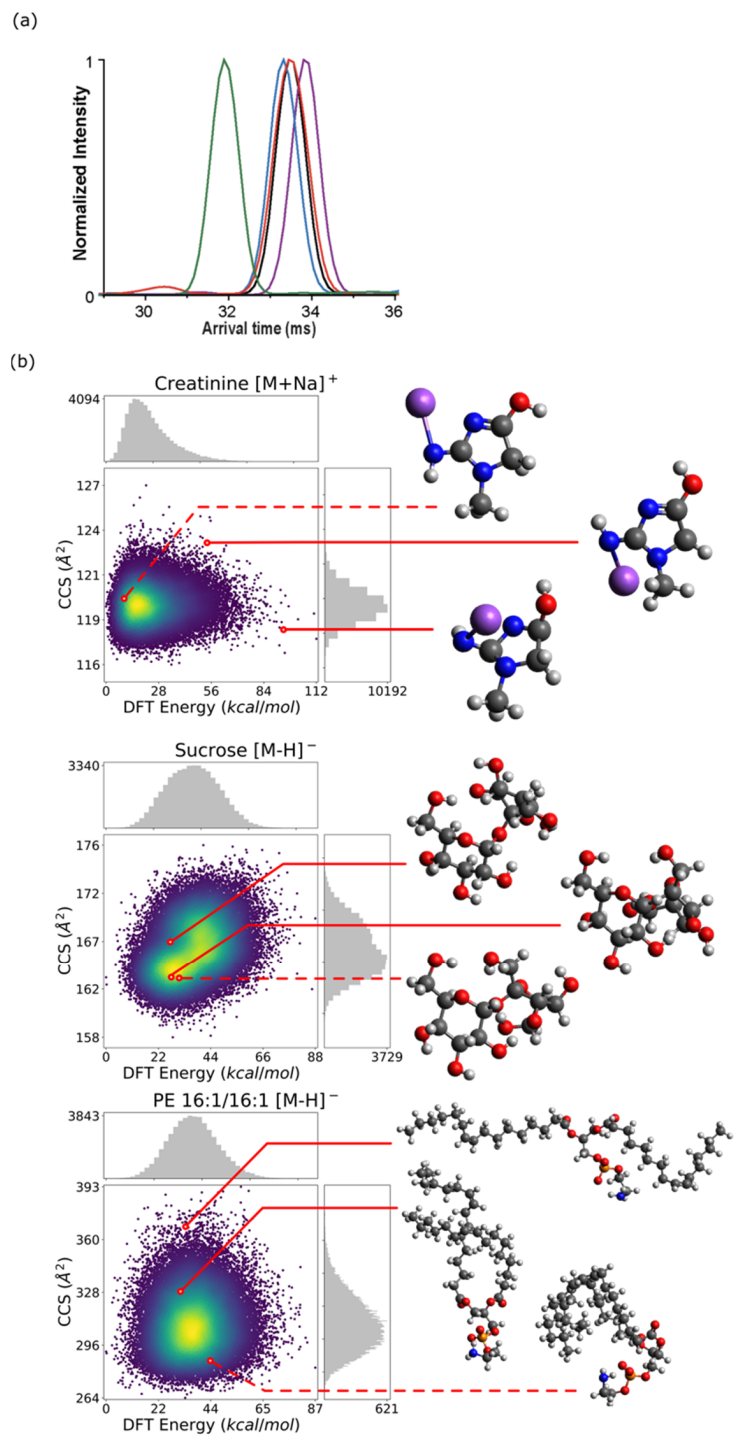

**Fig. 1 Example conformer-influenced empirical arrival time and *in silico* CCS distributions.** (a) Di-CQA isomers were shown by Zheng et al. to have overlapping distributions in DTIMS [33]. Distributions in IMS are believed to be largely due to diffusion and conformers. Reprinted (adapted) with permission from Zheng et al. Copyright (2017) American Chemical Society. (b) CCS vs energy landscapes for 50k AMBER generated conformers for creatinine [M+Na]$^+$, sucrose [M+Na]$^+$, and PE 16:1/16:1 [M-H]$^-$ respectively. Highlighted are the most similar and two most dissimilar conformers chosen heuristically with a structural RMSD metric.

**METHODS:**

*Benchmark Molecule Set*

We used a benchmark set of 18 small molecules reported in Colby et al. (2019) [32] with masses ranging from 113 to 687 Da. Experimental CCS values for benchmark set adducts ($[M+H]^+$, $[M-H]^+$, or $[M+Na]^+$) were obtained using an Agilent 6560 Ion Mobility Q-TOF MS (Agilent Technologies, Santa Clara) with nitrogen buffer gas, as described in Zheng et al. [35]. This adduct set was also processed through ISiCLE ("Standard" calculation methods) to create an initial predicted CCS baseline. **Fig. 2** plots the *m/z* vs CCS for the benchmark set molecules, and **Table S1** provides their *m/z*, adduct types, chemical classes, chemical formulae, and experimental CCS.

*Conformer Definition*

We define a conformer as returned by conformer generation tools: Each structure is a conformer, regardless of energy or energy minima. This is important in applications like IMS, where any valid structure can contribute to the CCS. This is in contrast to the IUPAC definition, where a conformer is only a structure that sits at the minimum of a potential energy well. [36] This latter definition makes no reference to transition state structures that, although fleeting, are present during experiments and impact measured properties such as CCS.

*Conformer Generation and Processing*

To test sampling methods on large sets of conformers, a modified ISiCLE pipeline was used to generate ~50,000 conformers for each adduct in the benchmark molecule set. Specifically, the AmberTools17 [37] simulated annealing MD tool, Sander, was used with simulated temperatures of 300 K to 600 K for 1000 annealing cycles, from which 50 conformers were randomly selected out of each cycle at the 300 K level. After conformer generation, CCS values for each conformer were calculated using MOBCAL-SHM [32], a shared-memory version of MOBCAL written in C and optimized for HPC resources, yielding 135× speed up over the original MOBCAL. For parameters, please see the **Supporting Information**. Finally, DFT energies were calculated for each conformer using NWChem [21], with B3LYP [38-40] exchange-correlation and 6-31G* basis set, via ISiCLE.

For three molecular adducts (mandelonitrile $[M+H]^+$, creatinine $[M+Na]^+$, and sucrose $[M-H]^-$), 25k-50k of their conformers were additionally geometry optimized using DFT in NWChem, with CCS subsequently calculated by MOBCAL-SHM to enable comparison of conformer selection techniques on quantum chemistry optimized structures.

To briefly compare simulated annealing against other conformer generation methods, 50k conformers were generated for one molecular adduct, mandelonitrile $[M+H]^+$ using RDKit (v 2019.03.1, rdkit.org), and the lowest energy conformer using GFN2-xTB force field was generated for each adduct in the set using the Conformer-Rotamer Ensemble Sampling Tool (CREST, v 2.7.1) [41]. RDKit randomly generates conformers using distance geometry, where constraints bound the minimum and maximum pairwise distances between any two atoms [42]. Under the iMTD-GC workflow, CREST uses a mixture of meta-dynamics (MTD), MD, z-matrix crossing, and other methods to iteratively search for low energy conformers and fill out their conformations by finding their rotamers (conformers in this case being understood under the IUPAC definition, i.e. a conformer is only the lowest energy structure of a potential energy well).

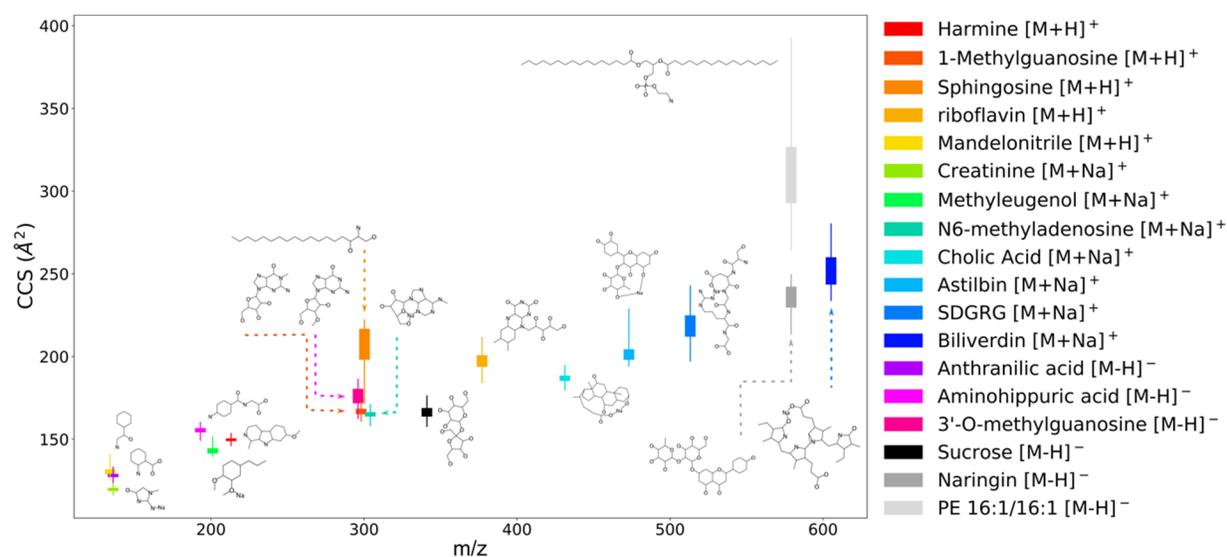

**Fig. 2** Ranges (thin line) and standard deviations (thick box) of CCS for a set of 18 small molecules.

*Conformer Geometry Variability*

Custom Python scripts (available for download on GitHub at https://github.com/PNNL/conformer_selection and provided in the **SI**) were created for performing Monte Carlo (MC) simulations in order to understand the root-mean-square deviation (RMSD) variability between conformer geometries as produced by simulated annealing. RMSD were calculated using the OpenBabel (v 2.4.1) OBAlign function [43,44] to align the conformers and calculate RMSD between corresponding heavy atoms (i.e. non-hydrogen atoms). The goal using MC was to simulate random draws from the true population of conformers for a range of sample sizes, where at each sample size, or step, conformers were randomly sampled and their RMSD averaged. Each MC simulation step was run for 10,000 iterations (see **Fig. S1** for justification) to produce a simulation average -- analogous to the most probable result when choosing a sample of that size -- and a standard deviation. Because simulated annealing works in cycles, we investigated two approaches when sampling: (1) The full 50k conformer set treated as a single pool (with a bias to sample across cycles) and (2) each cycle sampled as a group. This allowed for assessing possible correlations between adjacent cycles (**Fig. S2**). Details describing how sampling is applied in order to allow direct comparison between the two approaches is given in the **SI**.

Using Pearson product-moment correlation coefficient, three characteristics of the complete MC simulations, namely RMSD convergence point, final converged average RMSD value, and maximum standard deviation from the average RMSD, were correlated against 71 molecular properties calculated using ChemAxon's tool cxcalc [45], as well as against experimental CCS. The converged value is the final MC result when sampling the full population, and the convergence point is the sample size when the maximum standard deviation is within 0.01% of the converged value, as shown in **Fig. 4a**. We note the molecular property calculations were done on the parent (non-adduct) molecules and the MC convergence was measured on the ionized (adduct) molecules.

*Conformer Selection and CCS Averaging Methods*

Our goal was to sample from the full 50,000 conformer population of each adduct in order to simulate a situation in which a researcher had only generated the sampled conformers. The foundation for this decision was a hypothesis that the full 50,000 conformer population would represent the vast majority of the possible conformational space for the adducts in our benchmark molecule set. MC methods were used to simulate the result of CCS calculations after conformers were chosen from increasingly large conformer populations using a variety of selection techniques (described in more detail below): (1) simple average, (2) Boltzmann weighted average, (3) lowest energy, and (4) averaging below an energy threshold. These were chosen based on their prevalence in the literature [46-51].

In addition to these methods, a fifth technique, which preemptively down-selects from the full sample to the *m* most similar and *n* most dissimilar set of conformers, builds off of an approach introduced by Colby *et al.* [32], which provides a more computationally efficient method of sampling while maintaining high precision. This method and the previous four methods were used in tandem to analyze every possible combination of the methods for a set of parameters, as described in Results and Discussion, Section 5.

A schematic of the following selection techniques is shown in **Fig. 3**.

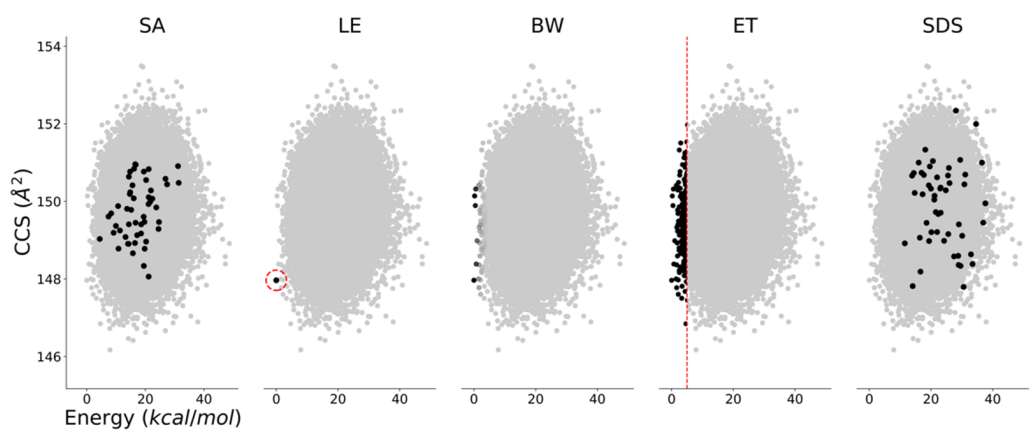

**Fig. 3 Diagram of conformer selection and downselection methods.** Simple average (SA), lowest energy (LE), Boltzmann weighting (BW), energy threshold (ET), and similarity downselection (SDS). SA shows 50 randomly selected conformers, LE shows the single lowest energy conformer, BW is shaded based on real weighted values, ET is a 5 *kcal/mol* threshold, and SDS shows the one most similar and 49 most dissimilar conformers.

*1. Simple average (SA)*

The simple average CCS is the arithmetic mean of all CCS values for a sample of conformers. Because the samples are randomly drawn from the full population, this simulates random conformer selection.

*2. Boltzmann Weighted (BW) average*

BW weights each conformer according to its Boltzmann probability distribution given by the equation,

$$p_i = \frac{e^{-\frac{E_i}{kT}}}{\sum_{i=1} e^{-\frac{E_i}{kT}}}$$

where $p_i$ is the probability or weight of the $i^{th}$ conformer, $E_i$ is the energy, $k$ is the Boltzmann constant, and T is temperature. Conformers that are lower in energy will have a higher weight when the CCS is averaged. This aims to reflect their time-averaged existence as a thermodynamic property and is heavily biased toward low energies. BW is currently considered the gold standard of conformer averaging used for CCS, NMR chemical shift calculations [47], as well as for other chemical properties.

*3. Lowest Energy (LE)*

Only the conformer with the single lowest energy is selected.

*4. Energy Threshold (ET)*

Only conformers with energy under the threshold are selected, and their CCS are simple averaged. Here, we apply 5, 2, 1, and 0.5 *kcal/mol* thresholds.

*5. Similarity downselection (SDS)*

The goal of SDS is to sample conformational space with fewer conformers while still being representative of the larger population. This can save on computational expense. Pairwise RMSD between conformations are used as a reciprocative similarity metric – the smaller the RMSD, the greater the similarity. SDS downselects based on this structural similarity to choose a subset of representative similar and most dissimilar conformers.

For this paper, we developed a heuristic algorithm for performing SDS and created an open source Python package that can be found at https://github.com/PNNL/SDS. The package includes relevant functions for performing SDS on conformers, but the SDS algorithm can also be generalized to any set of items where the items can be described as arrays whose elements are composed of the pairwise relations between the item in question and all other items of the set. Here, we employed the SDS algorithm to find the set of the *n* conformers most dissimilar from each other. To choose the most similar conformer, the pairwise RMSD between all conformers was summed, and the conformer with the smallest total RMSD was considered the most similar conformer.

MC simulation was run on BW, LE, SA, and ET (in combination with SA) at 1,000 iterations for each MC step. As with the RMSD analysis, the across- versus within-cycle approaches are also assessed here. MC analysis was done separately using AMBER potential energies and DFT energies.

**RESULTS AND DISCUSSION**

Our goal for this work was to evaluate many of the methods found in recent literature for sampling molecular conformations, especially with consideration for those methods that have been used for CCS calculations. Toward this end, we performed Monte Carlo analysis and various sampling techniques to assess conformational coverage (using RMSD) as well as the impact on CCS as a function of conformer sampling methods. This was done with a validation set of protonated, deprotonated, and sodiated adducts ([M+H]$^+$, [M-H]$^+$, [M+Na]$^+$) of various chemical classes spanning about 100-700 Da.

**Convergence of RMSD as a function of Monte Carlo sampled conformers**

MC simulations were run for 1,000 or 10,000 iterations (for CCS or RMSD analysis respectively) per data point (e.g. per number of conformers sampled) for each molecule in the validation set to ensure convergence. **Fig. 4** demonstrates an example MC convergence plot of the variability between conformer geometries as defined by average RMSD. The convergence plots of all molecules are almost indistinguishable when viewed separately (see **Fig. S3**). We note three characteristics of these plots distinct to each molecule: the final converged RMSD average at full population (which we refer to as the "converged value"), the maximum standard deviation, and the "convergence point", which we have defined as the sample size when the standard deviation reaches 0.01% of the final converged RMSD average.

As expected, a large converged value (a measure of the degree of variability between conformers for a molecule) is positively correlated with molecular mass ($r^2$: 0.67; p-value: < 1e-4), but also with properties such as chain atom/bond count ($r^2$: 0.91 / 0.93; p-value: < 1e-9 / < 1e-9), and rotatable bond count ($r^2$: 0.90; p-value: < 1e-08), and negatively correlated with the second acidic pKa site ($r^2$: 0.66; p-value: 1.3e-4). The converged value was also negatively correlated with having high ring counts; however, our data was insufficient for assessing statistical significance for these properties, and we would need to perform a study with a larger molecule set to verify this (**Fig. S5**). Note the positively correlated properties also correlate with molecular mass, reflecting how larger molecules typically have higher degrees of freedom than smaller molecules. A volcano plot showing the statistical significance and the magnitude of correlation for several other properties (**Fig. S6**) also revealed other properties such as the 3D Van der Waals surface area ($r^2$: 0.81; p-value: < 1e-6), and water accessible surface area ($r^2$: 0.80; p-value: < 1e-6) were highly correlated and significant. Also interesting are the convergence point results, because while a molecule may have a high converged value (high average pairwise RMSD of the entire 50,000 population), it could have a relatively small convergence point. In some cases, this may be because the variability of conformer space is sampled in relatively few conformers despite the large RMSD between those conformers. More correlations between MC convergence characteristics and molecular properties as a heatmap of Pearson r correlations for 71 properties are found in **Fig. S5**.

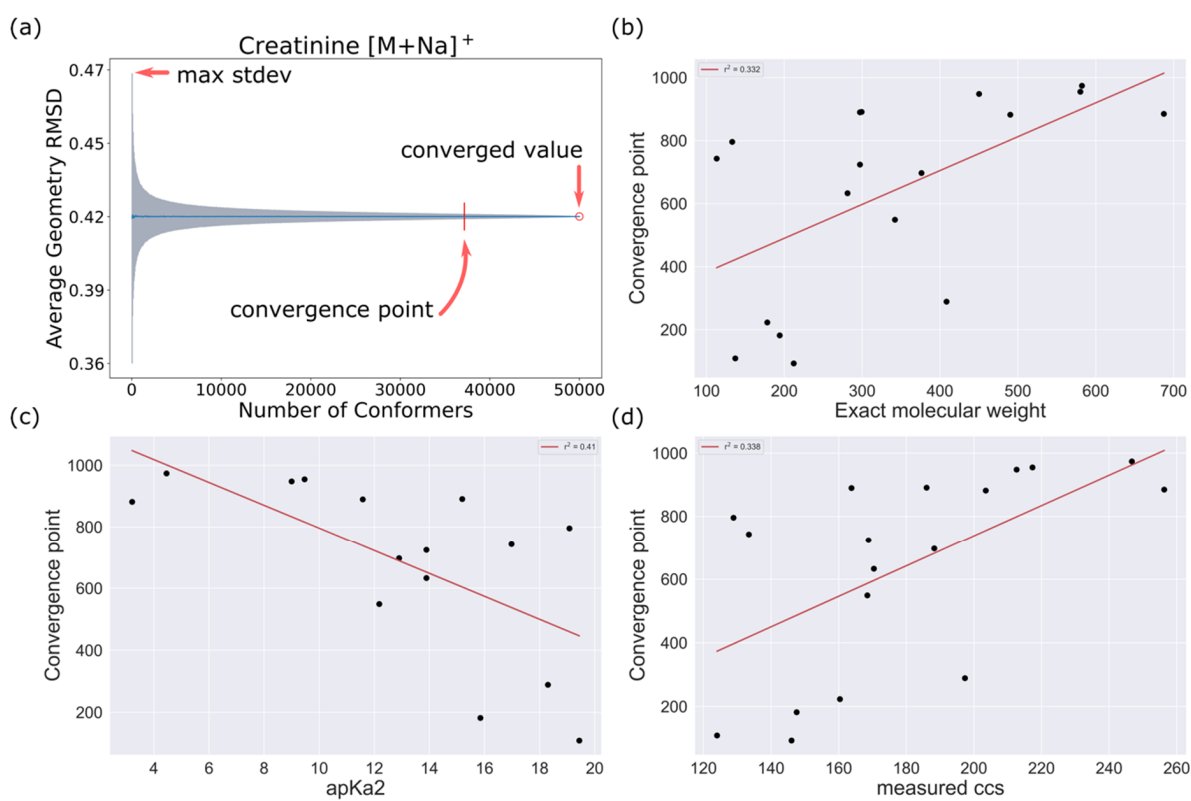

**Fig. 4 Example Monte Carlo simulation results on pairwise conformer RMSD.** (a) Example Monte Carlo convergence plot on RMSD between conformers for creatinine [M+Na]$^+$. Convergence point is the sample size (number of conformers) when standard deviation reaches 0.01% of the converged value. (b-d) Convergence point correlations with exact molecular weight, acidic pKa 2$^{nd}$ site (apKa 2), and experimentally measured CCS.

The Monte Carlo Sampling Methods section in the **SI** discusses additional details that consider convergence for within versus across simulated annealing cycles for molecular dynamics, revealing, as expected, that sampling across cycles resulted in better conformational space coverage. Interestingly, sampling within versus across cycles had little effect on the MC convergence of calculated CCS except to lower the precision of simple averaging methods when sampling within cycles, as seen in **Fig. S4.**

**Effect of conformer sampling on calculated CCS**

Ultimately, the desire is to assess the appropriate conformer sampling method producing stable and accurate chemical property predictions. In this manuscript, our application focused on CCS values. **Table S2** provides a summary of CCS calculated by each conformer sampling technique when all 50k conformers are sampled. The table also includes values from the best combination of these methods, ISiCLE [32], CREST [41], and experimental values. **Table 1** shows the mean absolute percent error relative to ISiCLE.

**Table 1 Mean absolute percent error of various conformer selection method results relative to ISiCLE CCS.** The ISiCLE method selects the most similar and two most dissimilar conformers out of 10 AMBER simulated annealing cycles (for a total of 30 conformers), applies DFT geometry optimization, and averages the CCS with Boltzmann weighting. Boltzmann weighting (BW), lowest energy (LE), simple average (SA), and simple averaging under energy thresholds 5, 2, 1, and 0.5 *kcal/mol* (ET 5, 2, 1, 0.5) were applied to 50k AMBER conformers and using DFT energies. CREST is the single lowest energy CREST conformer. The best combo is the statistical best combination we found on the AMBER conformers for the 18 molecules—10 AMBER cycles, selecting the most similar and 10 most dissimilar set under an AMBER energy threshold of 10 *kcal/mol,* and choosing the lowest energy according to the conformer's DFT energy.

| Molecule | BW | LE | SA | ET 5 | ET 2 | ET 1 | ET 0.5 | CREST | Best Combo | Experimental CCS |
|---:|---:|---:|---:|---:|---:|---:|---:|---:|---:|---:|
| Harmine +H | 0.33 | 0.48 | 0.70 | 0.49 | 0.27 | 0.45 | 0.44 | 0.35 | 0.66 | 1.78 |
| 1-Methylguanosine +H | 0.69 | 0.75 | 0.66 | 0.16 | 0.39 | 0.72 | 0.72 | 1.01 | 0.10 | 1.90 |
| Sphingosine +H | 11.75 | 14.83 | 14.85 | 13.47 | 12.32 | 12.30 | 8.95 | 5.36 | 2.13 | 2.99 |
| riboflavin +H | 0.10 | 0.10 | 2.26 | 0.40 | 0.51 | 0.69 | 0.69 | 5.46 | 0.23 | 2.42 |
| Mandelonitrile +H | 1.54 | 1.09 | 2.59 | 2.54 | 2.08 | 1.09 | 1.09 | 3.32 | 1.63 | 1.41 |
| Creatinine +Na | 0.41 | 0.56 | 0.66 | 0.35 | 0.34 | 0.45 | 0.65 | 0.17 | 0.67 | 12.24 |
| Methyleugenol +Na | 1.10 | 0.89 | 0.86 | 0.88 | 0.86 | 1.17 | 1.69 | 1.66 | 0.96 | 13.10 |
| N6-methyladenosine +Na | 0.51 | 1.22 | 0.54 | 0.17 | 0.14 | 0.40 | 0.77 | 0.66 | 0.32 | 2.66 |
| Cholic Acid +Na | 0.41 | 0.42 | 0.61 | 0.21 | 0.33 | 0.47 | 0.89 | 1.15 | 0.72 | 6.23 |
| Astilbin +Na | 2.76 | 3.68 | 0.14 | 0.88 | 1.55 | 1.55 | 3.68 | 0.94 | 1.17 | 5.83 |
| SDGRG +Na | 0.42 | 0.42 | 3.60 | 0.42 | 0.42 | 0.42 | 0.42 | 7.51 | 1.02 | 10.19 |
| Biliverdin +Na | 0.33 | 0.47 | 2.28 | 2.97 | 2.22 | 0.47 | 0.47 | 0.24 | 2.95 | 4.24 |
| Anthranilic acid -H | 0.07 | 0.09 | 0.16 | 0.03 | 0.16 | 0.23 | 0.16 | 0.89 | 0.80 | 3.02 |
| Aminohippuric acid -H | 1.66 | 1.70 | 2.17 | 1.66 | 1.61 | 1.70 | 1.70 | 0.11 | 3.25 | 3.02 |
| 3'-O-methylguanosine -H | 0.20 | 0.50 | 4.40 | 0.45 | 0.32 | 0.43 | 0.43 | 0.02 | 1.92 | 2.90 |
| Sucrose -H | 0.11 | 0.06 | 1.78 | 1.17 | 0.16 | 0.16 | 0.16 | 3.52 | 0.29 | 3.12 |
| Naringin -H | 0.86 | 1.14 | 2.07 | 0.86 | 0.15 | 1.14 | 1.14 | 8.95 | 2.56 | 9.77 |
| PE 16:1/16:1 -H | 6.06 | 6.07 | 0.51 | 1.97 | 6.24 | 6.24 | 6.24 | NaN | 1.04 | 17.67 |
| MAPE | 1.63 | 1.91 | 2.27 | 1.62 | 1.67 | 1.67 | 1.68 | 2.43 | 1.25 | 5.80 |

*1. SA, BW, and LE selection techniques on AMBER generated conformers*

**Fig. 5** demonstrates a convergence plot of the BW, LE, and SA sampling techniques. Consistently for all molecules, LE had the widest standard deviation (see **Fig. S7**). Both LE and BW had averages that sometimes skewed dramatically and had much higher standard deviations than the simple average. This happens whenever the conformer generation randomly produces a small population of one or more conformers with energies significantly lower than the rest. Because LE and BW are heavily biased toward lower energies, their selection was affected by the increasing probability of the MC simulation

"generating" lower energies as the sample size increased. A non-linear dependency between CCS and energy means the average CCS will keep changing as sample size increases to include more conformers with lower energies, and two samples of the same size may have widely different outcomes leading to high standard deviation. Therefore, the results of selection techniques dependent on energy are essentially functions of the CCS versus energy landscape. More specifically, they are functions of the low energy region, which is sparsely populated by AMBER. This sparsity leads to lower precision for low-energy dependent selection techniques. Thus, in order to understand how the selection techniques will behave, it is crucial to first understand how the conformer generation and optimization techniques shape the CCS (or other calculated value) versus energy landscape. Conformer generators like AMBER can produce significantly lower energy conformers even after thousands of generations (see **Table S4**), reducing precision for BW and LE between simulations.

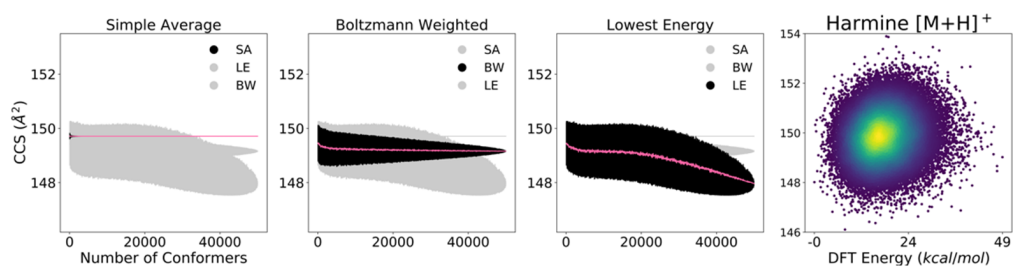

**Fig. 5 MC simulation convergence plots on CCS for harmine [M+H]$^+$.** Black and gray represent standard deviation from the average (pink). In this example, SA converges to the CCS where the conformers are most dense in the CCS vs energy landscape, BW averages between two low energy clusters, and LE converges to the single lowest energy CCS after 50k conformers are selected. This is a particularly good example showing the separation of the three techniques. Plots for all molecules of the set are shown in **Fig. S7**.

We note although the standard deviations of BW and LE seem wide relative to SA, it is misleading to think SA is the better option. SA may give the best precision in a computational model, but its choice does not reflect the underlying physics. For this reason, BW is recommended as it has better precision than LE and is still reflective of known physics. BW is supported in the literature as the current gold standard approach, even for other properties such as NMR chemical shift calculations [47,48,52]. In IMS in particular, a molecule does not exist as a single conformer, but rather interconverts between low-energy-barrier conformations rapidly during flight at room temperature. The final arrival time captured is then a weighted average of these conformers according to their duration of existence. It is reasonable to assume BW is more accurate because this is what is intended to be captured with Boltzmann's energy and temperature-dependent probability equation. At this time, however, we cannot directly say anything about the accuracy of the techniques with high confidence. Other underlying conditions, such as the ionization site location, can significantly influence conformer formation, changing the CCS versus energy landscape, and thus altering the predicted CCS. To achieve optimal accuracy, a wholistic approach needs to be taken, optimizing all aspects that could significantly change conformation simultaneously. Regardless, **Table 1** and **Table S3** have been provided as comparisons of the various selection techniques to the method implemented in ISiCLE (DFT geometry optimized conformers) and experimental values, respectively. Both tables suggest BW and LE have lower mean absolute percent errors, and therefore better accuracy, than SA.

See Section 5 for a discussion on the combination of multiple techniques.

*1.a SA, BW, and LE selection techniques on RDKit generated and DFT geometry optimized conformers*

A molecule's CCS vs energy space is shaped differently by different conformer generation and optimization methods, which can lead to significantly varied final calculated properties, even when using the same conformer selection techniques. **Fig. 6** compares the same selection techniques (SA, BW, and LE) as discussed above, but applied to RDKit generated conformers and DFT geometry optimized conformers (starting from AMBER generated structures). This is shown for 50k mandelonitrile [M+H]$^+$ conformers. Like AMBER, RDKit sparsely captured the low energy region of conformer space, leading to lower precision for LE. In this example, BW had a precision more comparable to SA's high precision, but this is likely an anomaly due to a split conformer population, since the low energy region is still sparsely populated. We note the RDKit conformers were not optimized using RDKit's universal force field (UFF) optimization tool. See **Fig. S8** for a discussion on how this may affect CCS.

DFT geometry optimization, on the other hand, significantly lowers the energy of all conformers and clusters them into "bars" where the energies are very close but the range of CCS remains wide (e.g. ~0.05 *kcal/mol* versus ~4 Å$^2$ for the example shown in **Fig. 6**). It appears that small changes, even the rotation of a methyl group on a rotamer, can lead to strikingly different CCS (e.g. one rotation by ~39 deg on creatinine's methyl group yielded a difference of ~1.1 Å$^2$, or ~0.95%). DFT geometry optimized conformers for creatinine [M+Na]$^+$ and sucrose [M-H]$^-$ are plotted in **Fig. S9**, showing similar results. DFT optimization densely populates the low energy region of CCS versus energy space, allowing BW and LE to have better precision. For example, the max standard deviation of BW for mandelonitrile dropped from σ 0.99 Å$^2$ to σ 0.08 Å$^2$, suggesting a few DFT geometry-optimized conformers are more effective than a large series of MD-based structures for small rigid molecules. How far this translates to larger, more flexible molecules is yet unknown. For sucrose, BW and LE had significant increases in precision, but the standard deviation of BW did not drop to 1% of the converged value until 650 conformers were randomly sampled, whereas mandelonitrile and creatinine were below 1% at the first sample size (50 conformers).

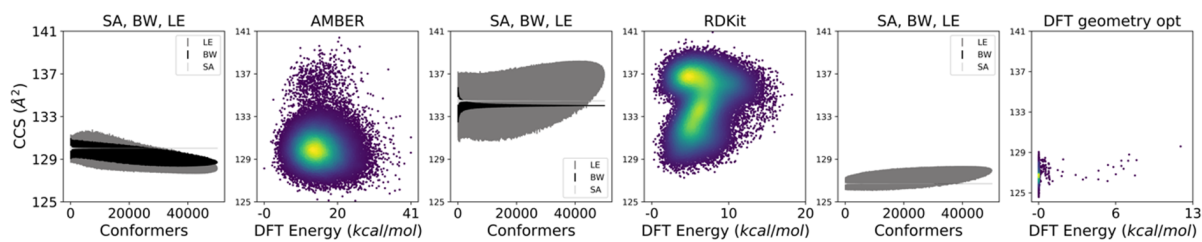

**Fig. 6 MC simulation convergence plots of CCS using three sampling techniques (SA, BW, LE) for conformers generated in AMBER, RDKit, and the AMBER conformers after a DFT geometry optimization for mandelonitrile [M+H]$^+$.** Interestingly, RDKit sampled a part of the CCS vs energy landscape that AMBER under sampled, and DFT geometry optimization collapsed the AMBER landscape into a single "bar" cluster where structures had similar energy, but subtle distinctions (e.g. rotamers) led to significantly different CCS. Under all three generation/optimization techniques, LE had the least precision. For all three molecules tested under DFT geometry optimization, BW precision improved dramatically. For this example, BW and SA happened to have the same effect after DFT geometry optimization (their convergence plots exactly overlap). This was not the case for the other two molecules tested.

*2. Averaging the CCS of conformers under energy thresholds*

Performing simple averages of all conformers under energy thresholds has historically been used as an alternative to choosing only the lowest energy conformer or performing Boltzmann weighting. **Fig. 7** compares example convergence plots of the ET method for 5, 2, 1, and 0.5 *kcal/mol* energy thresholds for mandelonitrile [M+H]$^+$. Plots for all molecules are included in **Fig. S10**. At certain thresholds, ET mimics the other selection techniques. This is no surprise because SA is the same as a threshold so large it encompasses every conformer, and LE is the same as a threshold so small it captures only one conformer. Thus, ET is bound by SA and LE methods. ET suffers from the same undersampling of the low energy region of CCS vs energy space that BW and LE do; as many as 1,051 or as few as 1 conformer were found for 5 *kcal/mol* threshold depending on the molecule, as shown in **Fig. S11**. The sparsity of the low energy region further complicates how to recommend which energy threshold is best, but there is a general trend that higher thresholds give higher precision between simulations at the expense of theoretical accuracy, and lower thresholds sacrifice precision when using MD conformers.

After selecting the conformers under an energy threshold, more than just SA can be applied. In Section 5, we describe a combination of methods where other methods, such as BW and SDS, are applied after the energy threshold reduction.

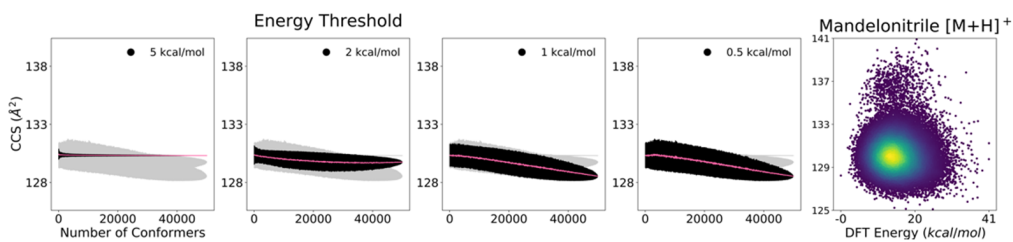

**Fig. 7 MC convergence plots on CCS for mandelonitrile [M+H]$^+$ for 5, 2, 1, and 0.5 *kcal/mol* energy thresholds.** Black represents standard deviation from the average (pink).

### 3. *RMSD based Similarity Downselection*

In our previous work, we found choosing the two most dissimilar conformers and the single most similar conformer from simulated annealing cycles (based on RMSD), yielded final CCS results within 99% of the result obtained from using all conformers from each cycle [32]. Building on this idea, we wanted to test how incrementally adding increasingly dissimilar conformers would impact the final property prediction. The idea behind SDS is to cover conformational space with fewer structures, thus maintaining accuracy while saving on computational expense. Ermanis et al. recently employed an RMSD-based similarity downselection method to exclude structures that were already very similar to each other and found using the 25 most dissimilar conformers was sufficient to minimize computational costs for NMR structure elucidation on small, rigid molecules [53]. We use SDS in an evaluation of different technique combinations as described below in Section 5.

### 4. *Using MD vs DFT energy on MD structures*

Conformers not at energy minima or in strict transition states are not well defined by quantum mechanical methods, and so calculating DFT energies on MD structures that have not been optimized by DFT are thought to be untrustworthy. However, we found DFT energy on MD-generated structures, before DFT geometry optimization, has better correlation to the CCS and energy of conformers after DFT geometry optimization (**Fig. 8**). Whereas MD energies have almost no correlation, DFT energies cluster the MD structure space in a way that can be mapped to the DFT geometry optimized space. If one can predict which cluster will map to the lowest energy DFT geometry optimized "bar," one can then select those structures to perform full DFT geometry optimization (assuming the goal is to get the lowest energy conformers at a given temperature). Since both MD and DFT energy calculations (on MD structures) run orders of magnitude faster than full DFT geometry optimization (see **Fig. S12** and **Table S5**), predicting beforehand which conformers to geometry optimize would result in considerable speed up. For molecules like creatinine, a low DFT energy threshold would suffice to secure conformers from this cluster. For sucrose, the cluster mapping to the lowest energy DFT geometry optimized "bar" was located closer to the middle of the MD "cloud," making it unclear how to successfully select those conformers without hindsight. Even so, there remains a general trend that higher energy clusters mapped to higher energy geometry-optimized "bars" and lower energies to lower geometry-optimized "bars" when using DFT energies.

While using DFT energies on non-optimized structures is not best practice in general, we feel our findings not only validate the use of DFT energies on MD structures for energy-based conformer space reduction methods such as energy thresholds, but also are better than using MD energies, especially with the goal of performing DFT geometry optimization on the structures afterwards. Analyses using MD energies, which showed similar trends between conformer selection techniques, were also done and can be found in the **SI**. Additionally, **Fig. S13** plots CCS vs energy space for all molecules, using AMBER energies, DFT energies, and DFT geometry optimized structures.

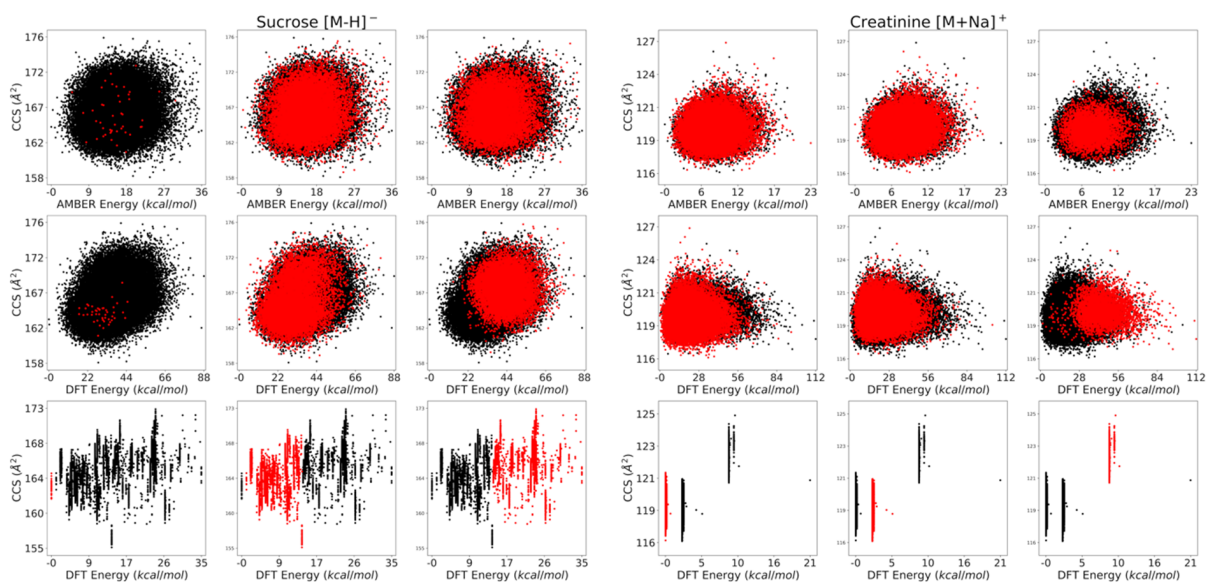

**Fig. 8 Demonstrations of clustering between DFT geometry optimized and non-optimized AMBER CCS vs energy space.** Specific clusters of conformers from DFT geometry optimized space (bottom) are chosen and highlighted in red. They are compared with the corresponding source conformers before optimization, using DFT energy (middle) and MD energy (top). DFT energy on MD structures has better correlation with the DFT geometry optimized structures than MD energy. DFT energy clearly predicts the fate of the conformers after DFT geometry optimization, whereas this is not evident with MD energy.

*5. Exploring the best combinations of space-reduction and selection techniques*

As a final exploration of conformer sampling techniques, we wanted to test combinations of the methods described above to see if there was a clear method or combination resulting in high accuracy, precision, and throughput. We ran an analysis exploring over 1,700 combinations using many different energy thresholds (both DFT and MD energies, alternatively or in combination), RMSD-based downselection (SDS), and conformer selection method combinations, including how many conformers were initially generated. The BW value of DFT geometry optimized conformers in the same manner as described in Colby et al. (2019) [32] was used to define the baseline in which to assess each combination. For sucrose, creatinine, and mandelonitrile, where we had 25k-50k DFT geometry optimized structures, all of their optimized conformers were used for the baseline. The method combination with the lowest mean absolute percent error (MAPE) was running AMBER for 10 cycles (50 conformers generated per cycle), using a 10 *kcal/mol* AMBER energy threshold followed by SDS to choose the 1 most similar and 10 most dissimilar, and then selecting the lowest energy conformer by DFT energy. This resulted in 1.2% (σ: 0.9%) MAPE and is estimated to take 200 (σ 157) minutes.

*6. Limitations of the study*

- The RMSD for SDS and geometry variability were calculated by excluding the non-backbone hydrogens. However, it appears even rotating a methyl group can lead to significantly different

- CCS calculations, and this appears to contribute to the CCS range of DFT geometry optimized conformers.
- Duplicate or nearly identical conformers generated by chance may give those conformers more weight than they should when Boltzmann weighting. This is a general problem for conformer generation tools like AMBER that don't screen for duplicate conformers. Tools like CREST check for this duplicity.
- In this analysis, we assume the lowest energy conformers for a given temperature are the conformers that would be present empirically. While this is usually accepted to be correct, in IMS, a molecule can become locked in a higher-energy local minima conformation

**CONCLUSION**

Using Monte Carlo analysis, we have shown the relative precision and behavior of various conformer selection techniques on AMBER generated conformers. Of the averaging or consolidation techniques (BW, LE, SA), BW had better precision than LE, and is physics-based, unlike SA, and is therefore expected to be more accurate. Example analysis on RDKit and DFT geometry optimized conformers confirms this trend, and also demonstrates the need for more efficient conformer generation tools that more thoroughly target the low energy region of conformer space. MD-based conformer generation tools like AMBER sparsely populate the low energy region of conformer space, leading to lower precision between simulations for energy-based selection methods, such as BW, LE, and ET. Applying robust structure optimization methods like DFT geometry optimization can help ameliorate this problem, greatly increasing the precision and expected accuracy of e.g. BW, but this comes at the cost of greater computational expense. For this reason, we have hopes for tools like CREST [41] and BOKEI [54] which use mixtures of methods (e.g. MTD, genetic z-matrix crossing, and semi-empirical methods) to specifically target low energy conformer space. A preliminary test of CREST showed it consistently generated conformers with lower (DFT) energies than AMBER for all molecules, half of the examples having energies as low as the DFT geometry optimized conformers. Subsequent research rigorously testing CREST against other methods is needed.

We ran a method-combinations search to find an optimal combination of the selection techniques for accuracy, precision, and computational expense when using non-DFT-optimized AMBER generated conformers. Of the conformer downselection techniques (ET, SDS, random), SDS and ET appeared to give lower MAPE than random methods. For single field experimental CCS methods, an interlaboratory study demonstrated an average uncertainty of 0.54%,[55] yet a recent study reported measured CCS uncertainty estimates of 4.7–9.1%.[56] For building *in silico* chemical libraries, we hope to achieve <1% MAE to meet future improvements of experimental platforms. The best method combination found had a MAPE greater than 1%. This further suggests the precursory conformer generation step was insufficient for our purposes. A more thorough analysis with a larger molecule set and tighter parameters would be needed to confirm this.

Many of our conclusions have been assumed in the literature, but here we have provided evidence for them. Already, researchers have been gravitating toward improved conformational sampling methods (e.g. those emerging from Prof. Stefan Grimme's group), showing a shift away from older, less relevant methods. In summary, we recommend Boltzmann weighting conformers as generated and optimized by tools that sufficiently populate the low energy region of conformer space (e.g. CREST and DFT geometry optimization). Doing so is expected to increase accuracy and precision while minimizing computational expense.

**ACKNOWLEDGMENTS**

This work was supported by the National Institutes of Health, National Institute of Environmental Health Sciences grant U2CES030170. Pacific Northwest National Laboratory (PNNL) is operated for the U.S. Department of Energy by Battelle Memorial Institute under contract DE-AC05-76RL01830.

# CITATIONS